\documentclass[12pt]{article}
\usepackage[latin1]{inputenc}\usepackage[english]{babel}
\usepackage{amsmath}
\usepackage{amsfonts}
\usepackage{latexsym}
\usepackage{url}

\newtheorem{theorem}{Theorem}[section]
\newtheorem{rem}[theorem]{Remark}
\newenvironment{remark}{\begin{rem}\rm}{\end{rem}}

\newtheorem{lemma}[theorem]{Lemma}

\newtheorem{eg}[theorem]{Example}

\newtheorem{conj}[theorem]{Conjecture}

\newtheorem{prob}[theorem]{Problem}

\newcommand{\pf}{\noindent {\bf Proof: }}

\begin{document}
\title{Fixed points avoiding Abelian $k$-powers}
\author{James D. Currie\thanks{The author is
supported by an NSERC Discovery Grant.}\\
Department of Mathematics and Statistics \\
University of Winnipeg \\
515 Portage Avenue \\
Winnipeg, Manitoba R3B 2E9 (Canada) \\\vspace{.1in}
\url{j.currie@uwinnipeg.ca} \\
Narad Rampersad\\
 Department of Mathematics \\
  University of Li{\`e}ge \\
  Grande Traverse 12 (B37) \\
  B-4000 Li{\`e}ge \\
  Belgium\\
\url {nrampersad@ulg.ac.be}
}
\maketitle
\begin{abstract}We show that the problem of whether the fixed point of a morphism avoids Abelian $k$-powers is decidable under rather general conditions.
\end{abstract}

\section{Introduction}
Let $\Sigma$ be a finite alphabet.  Consider the following decision
problem:
\begin{quote}
\it Given a morphism $h : \Sigma^* \to \Sigma^*$ with an infinite
fixed point $w$ and an integer $k \geq 2$, determine if $w$ is
$k$-power free.
\end{quote}
This decidability of this problem has been well studied.  For instance, Berstel \cite{berstel}
showed that over a ternary alphabet, there is an algorithm to
determine if $w$ is squarefree.  Similarly, Karhum\"aki \cite{karhumaki} showed that
over a binary alphabet, there is an algorithm to determine if $w$ is
cubefree.  The problem was solved in general by Mignosi and
S\'e\'ebold \cite{mig_see}, who showed that there exists an algorithm for this
problem for all alphabet sizes and all $k$.  (See also the work of
Krieger \cite{krieger} for extensions to fractional repetitions.)

In this paper we consider the analogous question for Abelian $k$-power
freeness.  In particular, we show that for morphisms $h$ satisfying
certain rather general conditions, the following problem is decidable.
\begin{quote}
\it Given a morphism $h : \Sigma^* \to \Sigma^*$ with an infinite
fixed point $w$ and an integer $k \geq 2$, determine if $w$ is
Abelian $k$-power free.
\end{quote}
Dekking \cite{dekking} provided sufficient conditions for a morphism $h$ to be
Abelian $k$-power free (i.e., $h$ maps Abelian $k$-power free words to
Abelian $k$-power free words).  Carpi \cite{carpi} showed the existence of an
algorithm to decide if a morphism satisfying certain technical
conditions is Abelian squarefree.  Our decision procedure for the
problem stated above is based on the idea of ``templates'' used in
\cite{aberkane1} and \cite{aberkane2}.  The same idea also appears in the recent work of
\cite{cassaigne}.

\section{Preliminaries}
We freely use the usual notations of combinatorics on words and formal language theory. (See for example \cite{hopcroft,lothaire}.) Fix positive integer $m$ and alphabet $\Sigma=\{1, 2,\ldots, m\}$. We use ${\mathbb Z}$ to denote the set of integers, and ${\mathbb Z}^n$ to
denote the set of $1\times n$ matrices (i.e. row vectors) with integer
entries.
For $u,v\in\Sigma^*$ we write $u\sim v$ if $u$ and $v$ are anagrams of each other, that is, if $|u|_a=|v|_a$ for all $a\in\Sigma$. We define the Parikh map $\psi:\Sigma^*\rightarrow{\mathbb Z}^m$ by
$$\psi(w)=[|w|_1,|w|_2,\ldots,|w|_{m}], w\in\Sigma^*.$$
In other words, $\psi(w)$ is a
row vector which counts the frequencies of $1,2
 \ldots, m$ in $w$. For $w,v\in \Sigma^*$ we have $w\sim v$ exactly when $\psi(w)=\psi(v)$.
Let $k$ be a positive integer. An {\bf Abelian $k$-power} is a non-empty word of the form $X_1X_2\cdots X_k$ where $X_i\sim X_{i+1}$, $1\le i\le k-1$.


Let a morphism  $\mu:\Sigma^*\rightarrow\Sigma^*$ be fixed.
It will be convenient and natural for us to make some assumptions on $\mu$:
\begin{eqnarray}
\mu(1)&=&1x,\mbox{ some }x\in\Sigma^+\\
|\mu(a)|&>&1,\mbox{ for all }a\in\Sigma
\end{eqnarray}
It follows that if $u\in\Sigma^*$, then $|u|\le |\mu(u)|/2.$

The {\bf frequency matrix} of $\mu$ is the $m\times m$ matrix $M$ such that $M_{i,j}=|\mu(i)|_{j}$. The $i^{th}$ row of $M$ is thus the Parikh vector of $\mu(i)$.
 For $w\in\Sigma^*$ we have
$$\psi(\mu(w))=\psi(w)M.$$
We will need some matrix theory. A standard reference is \cite[Chapter~5]{HJ85}. An induced norm on matrices of
${\mathbb R}_{m\times m}$ is given by
$$|M|=\sup_{v\in{\mathbb R^m}}\frac{|vM|}{|v|}$$
where $|v|$ is the usual Euclidean length of vector $v$. We make the additional restriction  on $\mu$ that $M$ is non-singular and  
that $|M^{-1}|<1.$

Let $$L=\{w:uwv\in \mu^n(1)\mbox{ for some positive integer }n,\mbox{
some words }u,v \}.$$ Thus $L$ is the set of factors of
the image of 1 under iteration of $\mu$. Language $L$ is closed under
$\mu$, and each word of $L$ is a factor of a word of $\mu(L)$. 
Let $N=\max_{a\in\Sigma}|\mu(a)|$. 
\begin{lemma}\label{parse} If $w\in L$ and $|w|\ge N-1$ then we can write $$w = A^{\prime\prime}\mu(b)C'$$ such that $A^{\prime\prime}$ is a (possibly empty) suffix of $\mu(a)$ and $C'$  is a (possibly empty) prefix of $\mu(c)$ for some $a,b,c\in\Sigma$.\end{lemma}
\pf The alternative is that $w$ is an interior factor of some word of $\mu(\Sigma)$, forcing $|w|\le N-2.\Box$
\section{Ancestors and $k$-templates}
Let $k$ be a positive integer. A {\bf $k$-template} is a ($2k$)-tuple $$t=[a_1,a_2,\ldots a_{k+1},d_1,d_2,\ldots, d_{k-1}]$$ where
the $a_i\in\{\epsilon,1,2,\ldots, m\}$ and the $d_i\in {\mathbb Z}^m$.
We say that a word $w$ {\bf realizes} $k$-template $t$ if a non-empty factor  ${\cal I}$ of $w$ has the form
$${\cal I}=a_1X_1a_2X_2a_3\ldots a_kX_ka_{k+1}$$ where $\psi(X_{i+1})-\psi(X_i)=d_i$, $i=1,2,\ldots k-1$.
Call ${\cal I}$ an {\bf instance} of $t$.
\begin{remark}
The particular $k$-template 
$$T_k=[\epsilon,\epsilon,\ldots, \epsilon, \overrightarrow{0}, \overrightarrow{0},\ldots, \overrightarrow{0}]$$
will be of interest.
Word  ${\cal I}$ is an instance of $T_k$ if and only if  
 ${\cal I}$ has the form
$${\cal I}=X_1X_2\ldots X_k$$ where $\psi(X_{i+1})=\psi(X_i)$, $i=1,2,\ldots k-1$;
in other words, if and only if  ${\cal I}$ is an Abelian $k$-power.
\end{remark}

Let $$t_1=[a_1,a_2,\ldots,a_{k+1},d_1,d_2,\ldots,d_{k-1}]$$ and $$t_2=[A_1,A_2,\ldots,A_{k+1},D_1,D_2,\ldots,D_{k-1}]$$ be
$k$-templates.
We say that $t_2$ is a {\bf parent} of $t_1$ if 
$$\mu(A_i)=a_i'a_ia_i^{\prime\prime}\mbox{ for some words }a_i',a_i^{\prime\prime}$$ while
\begin{equation}\label{image of parikh}\psi(a_{i+1}^{\prime\prime}a_{i+2}')-\psi(a_i^{\prime\prime}a_{i+1}')+D_iM=d_i,1\le i\le k.\end{equation}

\begin{lemma} (Parent Lemma)
Suppose that $w\in\Sigma^*$ realizes $t_2$. Then $\mu(w)$ realizes $t_1$. 
\end{lemma}
\pf Let $w$ contain the factor $${\cal I}=A_1Y_1A_2Y_2\cdots \cdots A_{k}Y_kA_{k+1}$$
where $\psi(Y_{i+1})-\psi(Y_i)=D_i$. For each $i$, write $\mu(A_i)=a_i'a_ia_i^{\prime\prime}$ and let $X_i=a_i^{\prime\prime}\mu(Y_i)a_{i+1}'.$ Then
$$\mu({\cal I})=a_1'{a_1}{X_1}{a_2}X_2\cdots X_{k-1}{a_k}{X_k}{a_{k+1}}a_{k+1}^{\prime\prime}$$
 and for each $i$, 
\begin{eqnarray*}
\psi(X_{i+1})-\psi(X_i)
&=&\psi(a_{i+1}^{\prime\prime}\mu(Y_{i+1})a_{i+2}')-\psi(a_{i}^{\prime\prime}\mu(Y_{i})a_{i+1}')\\
&=&\psi(a_{i+1}^{\prime\prime}a_{i+2}')-\psi(a_{i}^{\prime\prime}a_{i+1}')+\psi(\mu(Y_{i+1}))-\psi(\mu(Y_{i}))\\
&=&\psi(a_{i+1}^{\prime\prime}a_{i+2}')-\psi(a_{i}^{\prime\prime}a_{i+1}')+\left(\psi(Y_{i+1})-\psi(Y_{i})\right)M\\
&=&\psi(a_{i+1}^{\prime\prime}a_{i+2}')-\psi(a_{i}^{\prime\prime}a_{i+1}')+D_iM\\
&=&d_i,\mbox{ by }(\ref{image of parikh})\end{eqnarray*}
and $\mu(w)$ contains the instance $a_1X_1a_2X_2\cdots X_ka_{k+1}$ of $t_1.\Box$ 

\begin{lemma}
Given a $k$-template $t_1$, we may calculate all of its parents.
\end{lemma}
\pf The set
of candidates for the $A_i$ in a parent, and hence for the
$a_i',a_i^{\prime\prime}$ is finite, and may be searched
exhaustively. Since $M$ is non-singular, a choice of values for
$a_i',a_i^{\prime\prime}$, together with given values $d_i$,
determines the $D_i$ by (\ref{image of parikh}).$\Box$ 

\begin{remark} Note that not all computed
values for $D_i$ may be in ${\mathbb Z}^m$; some $k$-templates may
have no parents.\end{remark}

Rewriting  (\ref{image of parikh}), 
$$D_i=\left(d_i+\psi(a_{i}^{\prime\prime}a_{i+1}')-\psi(a_{i+1}^{\prime\prime}a_{i+2}')\right)M^{-1}.$$ 
Since the $a_i',a_i^{\prime\prime}$ are factors of words of $\mu(\Sigma)$, there are finitely many possibilities for $c= \psi(a_{i}^{\prime\prime}a_{i+1}')-\psi(a_{i+1}^{\prime\prime}a_{i+2}').$ Let $C$ be the (finite) set of possible values for $c$.

Let {\bf ancestor} be the transitive closure of the parent relation. The $D_i$ vectors in any ancestor of $k$-template $T_k$ will have the form

$$D_i=c_q M^{-q} + c_{q-1} M^{q-1} + \cdots + c_1 M^{-1} + c_0, \quad c_j\in C,j=0,\ldots,q.$$
Let $c^*=\max\{|c|:c\in C\}$  and let
$r = c^*/(1 - |M^{-1}|)$.  We have 
$$\begin{array}{lll}
&|D_i|\\
  = & |c_qM^{-q} + c_{q-1} M^{q-1} + \cdots + c_1 M^{-1} + c_0| \\
 \leq & |c_q M^{-q}| + |c_{q-1} M^{q-1}| + \cdots + |c_1 M^{-1}| + |c_0|&\text{(triangle inequality)} \\
 \leq & |c_q||M^{-q}| + |c_{q-1}||M^{q-1}| + \cdots + |c_1||M^{-1}|
+ |c_0| 
 & \text{(property of the induced norm)}\\
\leq & |c_q||M^{-1}|^q + |c_{q-1}||M^{-1}|^{q-1} + \cdots +
|c_1||M^{-1}| + |c_0|&\text{(submultiplicativity)} \\
 \leq & c^* |M^{-1}|^q + c^* |M^{-1}|^{q-1} + \cdots +
c^* |M^{-1}| + c^* \\
 \leq & c^* \sum_{i=0}^\infty |M^{-1}|^i &
\text{(since $|M^{-1}| < 1$)} \\
 = & \frac{c^*}{1 - |M^{-1}|}\\ =& r.
\end{array}$$

Thus, the $D_i$ lie within a ball of radius $r$ in $\mathbb{R}^n$.
It follows that there are only finitely many $D_i$'s in $\mathbb{Z}^n$.

\begin{lemma}\label{finite}
Template $T_k$ has finitely many ancestors.
\end{lemma}
\noindent {\bf Proof:} There are finitely many choices for the $A_i\in\{\epsilon,1,2,\ldots,m\}$ and the $D_i$ in any ancestor.$\Box$

Suppose that in $k$-template $t_1$ we have $\lfloor\max_i |d_i|\rfloor=\Delta$.
Let  ${\cal I}$ be an instance of $t_1$, 
$${\cal I}=a_1X_1a_2X_2a_3\ldots a_kX_ka_{k+1}$$ where $\psi(X_{i+1})-\psi(X_i)=d_i$, $i=1,2,\ldots k-1$.

If $i>j$, we have

\begin{eqnarray*}
|X_i|-|X_j|&=&\sum_{n=1}^m\left(|X_i|_n-|X_j|_n\right)\\
&=&\sum_{n=1}^m\sum_{q=1}^{i-j}\left(|X_{j+q}|_n-|X_{j+q-1}|_n\right)\\
&=&\sum_{n=1}^m\sum_{q=1}^{i-j}\left(\psi(X_{j+q})^{(n)}-\psi(X_{j+q-1})^{(n)}\right)\\
&\le& \sum_{n=1}^m\sum_{q=1}^{i-j}\Delta\\
&\le&mk\Delta
\end{eqnarray*}

This can be argued with the opposite inequality, showing in total that 
$$||X_i|-|X_j||\le mk\Delta.$$
If for some $i$ we have $|X_i|\le N-2$, then for $1\le j\le k$ we have $|X_j|\le N-2+mk\Delta$. The greatest possible length of ${\cal I}$ would then be $$|X_i|+\sum_{j=1}^{k+1}|a_j|+\sum_{j\ne i}|X_j|\le N-2 + k+1 + (k-2)(N-2+mk\Delta).$$ 
If $|{\cal I}|>N+k-1+(k-2)(N-2+mk\Delta)$ then for each $i$ we have $|X_i|>N-2$ and repeatedly using  Lemma~\ref{parse} we can write
\begin{eqnarray*}{\cal I}&=&a_1X_1a_2X_2a_3\ldots a_kX_ka_{k+1}\\
&=&a_1a_1^{\prime\prime}\mu(Y_1)a_2'a_2a_2^{\prime\prime}\mu(Y_2)a_3'\ldots a_k^{\prime\prime}\mu(Y_k)a_{k+1}'a_{k+1}
\end{eqnarray*}
 where ${\cal J}=A_1Y_1\cdots A_kY_kA_{k+1}$ is a factor of $\mu^\omega(1)$, $\mu(A_i)=a_i'a_ia_i^{\prime\prime}$ for each $i$ and $X_i=a_i^{\prime\prime}\mu(Y_i)a_{i+1}'$.
 It follows that parent $t_2$ of $t_1$ is realized by a factor of $\mu^\omega(1)$; moreover, instance ${\cal J}$ of $t_2$ satisfies $|{\cal J}|<|{\cal I}|/2.$

\begin{lemma}\label{inv_par}(Inverse Parent Lemma)
Suppose that ${\cal I}$ is a factor of $\mu^\omega(1)$ which is an instance of $t_1$, and $|{\cal I}|>N+k-1+(k-1)[N-2+mk\Delta]$. Then for some parent $t_2$ of $t_1$, $\mu^\omega(1)$ contains a factor ${\cal J}$ which is an instance of $t_2$, and such that $|{\cal J}|<|{\cal I}|.$
\end{lemma}

\section{Decidability}
\subsection{Main Theorem}
\begin{theorem}Let $\mu$ be a morphism on $\{1,2,\ldots, m\}$ and $M$ the frequency matrix of $\mu$. Suppose that
\begin{eqnarray*}
\mu(1)&=&1x,\mbox{ some }x\in\Sigma^+\\
|\mu(a)|&>&1,\mbox{ for all }a\in\Sigma\\
|M^{-1}|&<&1\end{eqnarray*}
and $M$ is non-singular.
 It is decidable whether $\mu^\omega(1)$ is Abelian $k$-power free.
\end{theorem}
\noindent {\bf Proof:} Calculate the set $T$ of ancestors of $T_k$. By
Lemma~\ref{finite} this set is finite.  Word  $\mu^\omega(1)$ contains an Abelian $k$-power iff an instance of one of these ancestors is a factor of $\mu^\omega(1)$. For each $t=[a_1,a_2,\ldots,a_{k+1},d_1,d_2,\ldots,d_{k-1}]\in T$, let $D_t=\{d_1,d_2,\ldots,d_{k-1}\}$. Let $D=\cup_{t\in T}D_t$, and let $\Delta=\lfloor\max_{d\in D}|d|\rfloor.$
As per Lemma~\ref{inv_par}, the shortest instance (if any) in $\mu^\omega(1)$ of a template of $T$ has length at most
$N+k-2+(k-2)(N-2+mk\Delta)$. We therefore generate all the factors of $\mu^\omega(1)$ of this length, and test whether any contains an instance of one of these ancestors.$\Box$
\section{Example}
In the case $m=3$, $k=3$, Dekking showed that the fixed point of $\mu$ contains no Abelian $3$-powers, where
\begin{eqnarray*}
\mu(1)&=&1123\\
\mu(2)&=&133\\
\mu(3)&=&223.
\end{eqnarray*}
His method of proof was elegant, but somewhat particular to his morphisms. 

Here $N=4$. We have
$$M=\left[\begin{array}{ccc}
2&1&1\\
1&0&2\\
0&2&1
\end{array}\right]$$
which is non-singular. A calculation with Lagrange multipliers shows that $|M|\approx 0.8589 < 1$. Applying the approach given in our Theorem, computing in the SAGE environment, we find that 
$T_3=[\epsilon,\epsilon,\epsilon,\epsilon,\overrightarrow{0},\overrightarrow{0}]$ has  1293 parents and no``grandparents". Since $T_3$ is an ancestor of itself, there are 1294 ancestors in $T$. Examining these ancestors, we find that $\Delta=2$. We therefore test the factors of $\mu^\omega(1)$ of length $N+k-2+(k-2)(N-2+mk\Delta)=25.$ None of these contains an instance of a template in $T$. This gives an alternate, mechanical proof of Dekking's result.

The matrices for both of Dekking's morphisms, for Pleasants' morphism and for Ker\"{a}nen's morphism\cite{dekking,pleasants,keranen} are invertible and satisfy $|M^{-1}|<1$ as well as our conditions (1) and (2). It follows that the results of these different authors could also be proved via the approach of the present paper.
\section{Future work}
Ker\"anen \cite{keranen,keranen2} has constructed an Abelian 2-power free quaternary word which is the fixed point of a cyclic 85-uniform morphism. His exhaustive searches have shown that this is the shortest cyclic uniform morphism which works. One would hope that much shorter, if less symmetric, morphisms exist. The result contained here suggests a new exhaustive search, considering shorter, not necessarily symmetric morphisms. The hope is that a better proof of the existence of infinite quaternary words avoiding Abelian 2-powers will result.

\end{document}